\documentclass[aps,prd,preprint,superscriptaddress,tightenlines,%
nofootinbib]{revtex4}
\newcommand{\PRE}[1]{{#1}}   % Use if preprint style

\usepackage{amsfonts,amsmath,amsthm,amssymb}
\usepackage{mathrsfs}
\usepackage{epsfig}
\usepackage{color}
\begin{document}

\newcommand{\hhat}[1]{\hat {\hat{#1}}}
\newcommand{\pslash}[1]{#1\llap{\sl/}}
\newcommand{\kslash}[1]{\rlap{\sl/}#1}
\newcommand{\qq}{$^\sharp$ }
% item that is questionable.
\newcommand{\rmb}{{\color{red}$^\bigstar$} }
% item that is for simple reading.
\newcommand{\fp}{{\color{blue}$^\heartsuit$} }
% item that contains proof usually for reference.
\newcommand{\qa}{$^\mho$ }
% item that contains answer to the question.
\newcommand{\rp}{{\color{magenta}$^\Re$} }
% reference point.
\newcommand{\lab}[1]{\hypertarget{lb:#1}}
% to create labels.
\newcommand{\iref}[2]{\footnote{\hyperlink{lb:#1}{\textit{$^\spadesuit$#2}}}}
% reference to other part in this notes.
\newcommand{\emp}[1]{{\bf\color{red} #1}}
% Emphasis
\newcommand{\eml}[1]{{\bf \color{blue} #1}}
% Quick Guide.
\newcommand{\kw}[1]{\emph{#1}}
% Key word.
\newcommand{\sos}[1]{{\large \textbf{#1}}}
% Subsection
% \newcommand{\soso}[1]{{\large \begin{center} \textbf{#1} \end{center}}}
\newcommand{\soso}[1]{\chapter{#1}}
% Subsection ver 2. Maybe used to do more. One can change the "\[" part to use other like "\center" to recover the "\[" for equation number.
% Now is chapter
%\newcommand{\sossub}[1]{\[\textbf{#1}\]}
\newcommand{\sossub}[1]{\section{#1}}
% SubSubsection
\newcommand{\et}[1]{{\large\begin{flushleft} \color{blue}\textbf{#1} \end{flushleft}}}
% \newcommand{\et}[1]{{\color{blue}\textbf{#1}}\\}
% Entry
\newcommand{\sto}[1]{\begin{center} \textit{#1} \end{center}}
% special topic
\newcommand{\rf}[1]{{\color{blue}[\textit{#1}]}}
\newcommand{\set}[1]{\mathbb{#1}}
% Reference

\newcommand{\postscript}[2]{\setlength{\epsfxsize}{#2\hsize}
   \centerline{\epsfbox{#1}}}

%%%%%%%%%%%%%%%%%%%%%
\newcommand{\el}[1]{\label{#1}}
% Equation labeling
\newcommand{\er}[1]{\eqref{#1}}
% Equation Reference
\newcommand{\df}[1]{\textbf{#1}}
% Temporarily replace \textbf
\newcommand{\mdf}[1]{\pmb{#1}}
% Use for vectors etc.
\newcommand{\ft}[1]{\footnote{#1}}
% footnote.
\newcommand{\n}[1]{$#1$}
% Use for numbers etc.
% \newcommand{\cjktext}[1]{\begin{CJK}{GB}{gbsn} #1 \end{CJK}}
% Language support
\newcommand{\fals}[1]{$^\times$ #1}
% wrong statement
\newcommand{\new}{{\color{red}$^{NEW}$ }}
% update
% \newcommand{\ci}[1]{\cite{#1}}
\newcommand{\ci}[1]{}
% citation

\newcommand{\de}[1]{{\color{green}\underline{#1}}}
\newcommand{\ke}{\rangle}
\newcommand{\br}{\langle}
\newcommand{\lb}{\left(}
\newcommand{\rb}{\right)}
\newcommand{\blb}{\Big(}
\newcommand{\brb}{\Big)}
\newcommand{\nn}{\nonumber \\}
\newcommand{\p}{\partial}
\newcommand{\pd}[1]{\frac {\partial} {\partial #1}}
\newcommand{\cd}{\nabla}
\newcommand{\cc}{$>$}
% ##### ###### ###### ######
\newcommand{\ba}{\begin{eqnarray}}
\newcommand{\ea}{\end{eqnarray}}
\newcommand{\be}{\begin{equation}}
\newcommand{\ee}{\end{equation}}
\newcommand{\bay}[1]{\left(\begin{array}{#1}}
\newcommand{\eay}{\end{array}\right)}
\newcommand{\eg}{\textit{e.g.} }
\newcommand{\ie}{\textit{i.e.}, }
\newcommand{\iv}[1]{{#1}^{-1}}
\newcommand{\st}[1]{|#1\ke}
\newcommand{\at}[1]{{\Big|}_{#1}}
\newcommand{\zt}[1]{\texttt{#1}}
\newcommand{\zi}[1]{\textit{#1}}
% ##### ###### ###### ######
% Greek Letters
\def\xa{{\alpha}}
\def\xA{{\Alpha}}
\def\xb{{\beta}}
\def\xB{{\Beta}}
\def\xd{{\delta}}
\def\xD{{\Delta}}
\def\xe{{\epsilon}}
\def\xE{{\Epsilon}}
\def\xve{{\varepsilon}}
\def\xg{{\gamma}}
\def\xG{{\Gamma}}
\def\xk{{\kappa}}
\def\xK{{\Kappa}}
\def\xl{{\lambda}}
\def\xL{{\Lambda}}
\def\xo{{\omega}}
\def\xO{{\Omega}}
\def\xvp{{\varphi}}
\def\xs{{\sigma}}
\def\xS{{\Sigma}}
\def\xt{{\theta}}
\def\xT{{\Theta}}
% ##### ###### ###### ######
\def \Tr {{\rm Tr}}
\def\CA{{\cal A}}
\def\CC{{\cal C}}
\def\CD{{\cal D}}
\def\CE{{\cal E}}
\def\CF{{\cal F}}
\def\CH{{\cal H}}
\def\CJ{{\cal J}}
\def\CK{{\cal K}}
\def\CL{{\cal L}}
\def\CM{{\cal M}}
\def\CN{{\cal N}}
\def\CO{{\cal O}}
\def\CP{{\cal P}}
\def\CQ{{\cal Q}}
\def\CR{{\cal R}}
\def\CS{{\cal S}}
\def\CT{{\cal T}}
\def\CV{{\cal V}}
\def\CW{{\cal W}}
\def\CY{{\cal Y}}
\def\CZ{{\cal Z}}
\def\BC{\mathbb{C}}
\def\BR{\mathbb{R}}
\def\BZ{\mathbb{Z}}
\def\sA{\mathscr{A}}
\def\sB{\mathscr{B}}
\def\sD{\mathscr{D}}
\def\sE{\mathscr{E}}
\def\sF{\mathscr{F}}
\def\sG{\mathscr{G}}
\def\sH{\mathscr{H}}
\def\sJ{\mathscr{J}}
\def\sL{\mathscr{L}}
\def\sM{\mathscr{M}}
\def\sN{\mathscr{N}}
\def\sO{\mathscr{O}}
\def\sP{\mathscr{P}}
\def\sR{\mathscr{R}}
\def\sQ{\mathscr{Q}}
\def\sX{\mathscr{X}}

\def\slz{SL(2,\BZ)}
\def\slr{$SL(2,R)\times SL(2,R)$ }
\def\ads{${AdS}_5\times {S}^5$ }
\def\adst{${AdS}_3$ }
\def\sun{SU(N)}
\def\ad#1#2{{\frac \delta {\delta\sigma^{#1}} (#2)}}
% for SU(N) SYM
\def\bqf{\bar Q_{\bar f}}
\def\nf{N_f}
\def\sunf{SU(N_f)}

\def\dcirc{{^\circ_\circ}}

\def\btr{{B_\nu{}^\nu}}
\def\byy{{B_{yy}}}
\def\sy{{\zt{sgn}(y)}}

\title{LHC Phenomenology of  Lowest Massive Regge Recurrences\\
 in the Randall-Sundrum Orbifold
%\PRE{\vspace*{1.5in}}
\PRE{\vspace*{0.3in}} }

\author{Luis A. Anchordoqui}
\affiliation{Department of Physics,\\
University of Wisconsin-Milwaukee,
 Milwaukee, WI 53201, USA
\PRE{\vspace*{.1in}}
}

\author{Haim Goldberg}
\affiliation{Department of Physics,\\
Northeastern University, Boston, MA 02115, USA
\PRE{\vspace*{.1in}}
}

\author{Xing Huang}
\affiliation{Department of Physics,\\
University of Wisconsin-Milwaukee,
 Milwaukee, WI 53201, USA
\PRE{\vspace*{.1in}}
}

\author{Tomasz R. Taylor}
\affiliation{Department of Physics,\\
Northeastern University, Boston, MA 02115, USA
\PRE{\vspace*{.1in}}
}

\date{June 2010}
\PRE{\vspace*{.6in}}
\begin{abstract}\vskip 3mm
  \noindent We consider string realizations of the Randall-Sundrum
  effective theory for electroweak symmetry breaking and explore the
  search for the lowest massive Regge excitation of the gluon
  and of the extra (color singlet) gauge boson  inherent of D-brane
  constructions. In these curved backgrounds, the higher-spin Regge
  recurrences of Standard Model fields localized near the IR brane are
  warped down to close to the TeV range and hence can be produced at
  collider experiments. Assuming that the theory is weakly coupled, we
  make use of four gauge boson amplitudes evaluated near the first
  Regge pole to determine the discovery potential of LHC. We study the
  inclusive dijet mass spectrum in the central rapidity region
  $|y_{\rm jet}| < 1.0$ for dijet masses $M\geq 2.5~{\rm TeV}$. We
  find that with an integrated luminosity of 100~fb$^{-1}$, the
  5$\sigma$ discovery reach can be as high as 4.7~TeV. Observations of
  resonant structures in $pp\rightarrow {\rm direct}\ \gamma~ +$ jet
  can provide interesting corroboration for string physics up to
  3.0~TeV. We also study the ratio of dijet mass spectra at small and
  large scattering angles. We show that with the first~fb$^{-1}$ such
  a ratio can probe lowest-lying Regge states for masses $\sim
  2.5$~TeV.
\end{abstract}

\maketitle

\section{Introduction}

The saga of the Standard Model (SM) is still exhilarating because it
leaves all questions of consequence unanswered. Perhaps the most
evident of unanswered questions is why the weak interactions are weak.
The well-known non-zero vacuum expectation value of the scalar Higgs
doublet condensate, $\langle H \rangle = v = 2.46 \times 10^2~{\rm
  GeV}$, sets the scale of electroweak interactions. However, due to
the quadratic sensitivity of the Higgs mass to quantum corrections
from an arbitrarily high mass scale, one is faced with the gauge
hierarchy problem: the question of why $v \ll M_{\rm Pl}$, where
$M_{\rm Pl} = 1.22 \times 10^{19}~{\rm GeV}$ is the Planck
mass. The traditional view is to adopt $M_{\rm Pl}$ as  the
fundamental mass setting the scale of the unified theory
incorporating gravity and attempt to derive $v$ through some dynamical
mechanism (e.g. renormalization group evolution).  In recent years,
however, a new framework with a diametrically opposite viewpoint has
been proposed, in which $v$ is instead the fundamental scale of
nature~\cite{ArkaniHamed:1998rs}. D-brane string compactifications
with low string scale and large extra dimensions allow a definite
representation of this innovative premise~\cite{Antoniadis:1998ig}.

TeV-scale superstring theory provides a brane-world description of the
SM, which is localized on membranes extending in $p+3$ spatial
dimensions, the so-called D-branes~\cite{Blumenhagen:2006ci}. Gauge
interactions emerge as excitations of open strings with endpoints
attached on the D-branes, whereas gravitational interactions are
described by closed strings that can propagate in all nine spatial
dimensions of string theory (these comprise parallel dimensions
extended along the $(p+3)$-branes and transverse dimensions). The
apparent weakness of gravity at energies below a few TeV can then be
understood as a consequence of the gravitational force ``leaking''
into the transverse large compact dimensions of spacetime. This is possible
only if the intrinsic scale of string excitations is also of order a
few TeV. Should nature be so cooperative, a whole tower of infinite
string excitations will open up at this low mass threshold, and new
particles of spin $J$ follow the well known Regge trajectories of
vibrating strings: $J = J_0 + \alpha' M^2$, where $\alpha'$ is the
Regge slope parameter that determines the fundamental string mass
scale
\begin{equation}
M_s={1\over \sqrt{\alpha'}}\, .
\label{Ms}
\end{equation}
Only one assumption is necessary to build up a solid framework: the
string coupling must be small for the validity of perturbation theory
in the computations of scattering amplitudes. In this case, black hole
production and other strong gravity effects occur at energies above
the string scale, therefore at least the few lowest Regge recurrences
are available for examination, free from interference with some
complex quantum gravitational phenomena.

In proton collisions at the
Large Hadron Collider (LHC), Regge states will be produced as soon as
the energies of some partonic subprocesses cross the threshold at
$\hat s > M_s^2.$ In a series of recent
publications~\cite{Anchordoqui:2007da,Anchordoqui:2008hi,Lust:2008qc,Anchordoqui:2008di,Lust:2009pz,Anchordoqui:2009ja}
we have computed open string scattering amplitudes in D-brane models
and have discussed the associated phenomenological aspects of low mass
string Regge recurrences related to experimental searches for physics
beyond the SM~\cite{Anchordoqui:2009bn}. We developed our
program in the simplest way, by working within the construct of a
minimal model in which we considered scattering processes which take
place on the (color) $U(3)_a$ stack of D-branes, which is intersected
by the (weak doublet) $U(2)_b$ stack of D-branes, as well by a third
(weak singlet) $U(1)_c$ stack of D-brane. In
the bosonic sector, the open strings terminating on the $U(3)_a$ stack
contain the standard gluons $g$ and an additional $U(1)_a$ gauge boson
$C$; on the $U(2)_b$ stacks the open strings correspond to the weak
gauge bosons $W$, and again an additional $U(1)_b$ gauge field.  So
the associated gauge groups for these stacks are $SU(3)_C \times
U(1)_a,$ $SU(2)_L \times U(1)_b$, and $U(1)_c$, respectively;
the physical hypercharge is a linear combination of $U(1)_a,$
$U(1)_b$, and $U(1)_c$.  The fermionic matter consists of open
strings, which stretch between different stacks of D(p+3)-branes and
are hence located at the intersection points.

In canonical D-brane constructions the large hierarchy between the
weak scale and the fundamental scale of gravity is eliminated through
the large volume of the transverse dimensions. An alternative
explanation to solve the gauge hierarchy problem was suggested by
Randall and Sundrum (herein RS)~\cite{Randall:1999ee}. The RS set-up
has the shape of a gravitational condenser: two branes, which rigidly
reside at $S^1/\set{Z}_2$ orbifold fixed point boundaries $y = 0$ and
$y = \pi r_c$ (the UV and IR branes, respectively), gravitationally
repel each other and are stabilized by a slab of anti-de Sitter $AdS$
space. The metric satisfying this Ansatz (in horospherical
coordinates) is given by
\begin{equation}
ds^2 = e^{-2  k |y|} \,\,\eta_{\mu\nu} \,dx^\mu dx^\nu + dy^2\,,
\label{lisa-metric}
\end{equation}
where $ k$ is the $AdS$ curvature scale, which is somewhat smaller
than the fundamental 5-dimensional Planck mass $M^\star_{\rm Pl} \sim
M_{\rm Pl}$.\footnote{Greek subscripts extend over ordinary
  4-dimensional spacetime and are raised and lowered with the flat
  Minkowskian metric $\eta_{\mu \nu}$, whereas Latin subscripts span
  the full 5-dimensional space and are raised and lowered with the
  full metric $g_{MN}$.}  In this set up the distance scales get
exponentially redshifted as one moves from the UV brane towards the IR
brane. Such exponential suppression can then naturally explain why the
observed physical scales are so much smaller than the Planck scale.
For example, if the 5-dimensional Higgs condensate $v_5 \sim k$ is
IR-localized, the observed 4-dimensional value will be obtained from
$e^{-k\pi r_c} \langle H_5 \rangle$, and the observed hierarchy
between the gravitational and electroweak mass scales is reproduced if
$kr_c \approx 12$.  The most distinct signal of this set-up is the
appearance of a tower of spin-2 resonances, corresponding to the
Kaluza-Klein (KK) states of the 5-dimensional graviton, which have
masses and couplings driven by the TeV-scale. These KK gravitons
couple to all SM fields universally, yielding striking predictions for
collider experiments~\cite{Davoudiasl:1999jd}.

As originally noted in~\cite{Goldberger:1999wh}, to address the
hierarchy problem it is sufficient to keep the Higgs near the IR
brane.  Interestingly, if the remaining gauge bosons and fermions are
allowed to propagate into the warped  dimension, one can also
formulate an attractive mechanism to explain the flavor mass
hierarchy~\cite{Grossman:1999ra}.  The idea here is that the light
fermions are localized near the UV brane. This raises the effective
cutoff scale for operators composed of these fields far above the
TeV-regime, providing an efficient mechanism to suppress unwanted
operators, such as those mediating flavor changing  neutral currents
(FCNC) processes, related to tightly constrained light
flavors. Moreover, this results in small 4-dimensional Yukawa
couplings to the Higgs, even if there are no small 5-dimensional
Yukawa couplings. The top quark is IR-localized to obtain a large
4-dimensional top Yukawa coupling. Because the fermion profiles depend
exponentially on the bulk masses, this provides an understanding of
the hierarchy of fermion masses (and mixing) without hierarchies in
the fundamental 5-dimensional parameters, solving the SM flavor
puzzle.

The RS set-up has also been used to construct warped Higgsless models,
where the electroweak symmetry is broken by boundary conditions on the
5-dimensional gauge fields~\cite{Csaki:2003zu}. Gauge fields are
allowed to propagate within all 5 dimensions. The electroweak gauge
structure of the minimal viable model is $SU(2)_L \times SU(2)_R
\times U(1)_{B-L},$ where $U(1)_{B-L}$ corresponds to gauging baryon
minus lepton number. Boundary conditions on the bulk gauge fields are
chosen so that the $SU(2)_L \times SU(2)_R$  symmetry is broken
on the IR brane to the diagonal subgroup $SU(2)_D$, and the $SU(2)_R
\times U(1)_{B-L}$ symmetry is broken down to the usual $U(1)_Y$
hypercharge in the UV brane to ensure that the low-energy gauge group
without electroweak symmetry breaking is $SU(2)_L \times U(1)_Y$.
The $SU(3)_C$ QCD group is unbroken everywhere, i.e., in the warped
dimension and on the branes. The spectrum of electroweak vector bosons
consists of a single massless photon along with KK towers of charged
$W_n$ and neutral $Z_n$ states. The SM massive $W$ and $Z$ vectors,
which get masses from the $SU(2)_L \times U(1)_Y$-violating boundary
condition on the IR brane, are identified with the lowest KK modes of
the $W_n$ and $Z_n$ towers.  SM fermions extend into all dimensions,
and they have explicit mass terms that are allowed by the non-chiral
structure of the theory in the bulk and on the IR brane. The most
serious challenge to construct viable models of Higgsless electroweak
symmetry breaking is satisfying the constraints from precision
electroweak measurements~\cite{Nomura:2003du}.  Mixing of the $W$ and
$Z$ with higher KK modes changes their couplings to fermions relative
to the SM. Heavier KK modes ($m_{\rm KK} \agt 1~{\rm TeV}$) are
preferred to reduce these deviations to an acceptable level, but the
KK modes cannot be too heavy ($m_{\rm KK} \alt 1~{\rm TeV}$) if they
are to unitarize vector boson scattering. Both requirements can be
satisfied simultaneously if there are localized kinetic terms on each
of the branes~\cite{Cacciapaglia:2004jz}, and if the SM fermions (with
the exception of the right-handed top quark) have approximately flat
profiles in the extra dimension~\cite{Cacciapaglia:2004rb}.  In this
case, the first vector boson KK modes above the $Z$ and $W$ typically
have masses $\approx 0.5-1.5~{\rm TeV}$~\cite{Cacciapaglia:2004rb}.
Additional model structure is needed to generate a sufficiently large
top quark mass while not overly disrupting the measured $Z b_L \bar
b_L$ coupling. Some examples include new top-like custodial bulk
fermions~\cite{Agashe:2006at}, or a second warped bulk space on the
other side of the UV brane with its own IR
brane~\cite{Cacciapaglia:2005pa}.

In string realizations of extended RS models of hierarchy and flavor,
we do expect the higher-spin Regge recurrences of the SM fields
localized near the IR brane to be redshifted close to the TeV scale
and therefore be directly produced at the
LHC~\cite{Hassanain:2009at,Perelstein:2009qi}. In this paper, we
explore the search for the lowest massive Regge excitation in
inclusive $\gamma$ + jet and dijet mass spectra. In the spirit
of~\cite{Hassanain:2009at,Perelstein:2009qi,Reece:2010xj}, we {\it
  assume} that the RS orbifold arises as part of the compactification
manifold in a weakly-coupled string theory.  We further {\it assume}
that the compactification radii of the other five dimensions are
${\cal O}( {M_s^\star}^{-1})$ and therefore can be safely integrated
out.\footnote{The dearth of string constructions for a transition to
  the RS compactification~\cite{Verlinde:1999fy} makes a full
  comparison between the string scale and internal dimension radii
  difficult. A recent study~\cite{Reece:2010xj} of a range of models
  seems to indicate that $M_s^\star r_c \sim 1$ is viable.} With this
in mind, the basic relation between the curvature of the warped
internal space, the string scale (the mass of the Regge states), and
the 5-dimensional Planck mass is
 \begin{equation}   k \ll M^\star_s= \frac{1}{
    \sqrt{{\alpha'}^\star}} \ll M_{\rm Pl}^\star\,  ,
\label{Mstar}
\end{equation}
where ${\alpha'}^\star$ is the slope of the associated Regge
trajectory. The first inequality permits the warping to leave intact the
basic string properties (such as the dual resonant structure) of the perturbative
scattering amplitudes.
 The infinite tower of open string Regge
excitations have the same quantum numbers under the SM gauge group as
the gluons and the quarks, but in general higher spins, and their
masses are just square-root-of-integer multiples of the string mass
$M_s^\star$.  In what follows, the first Regge excitations of the gluon $(g)$, the
extra $U(1)$ boson tied to the color stack ($C$), and quarks $(q)$
will be indicated with $g^*,C^*, q^*$, respectively.  In this paper we
complement model independent searches of top-production via $q^*$
excitation~\cite{Hassanain:2009at,Dong:2010jt} by analyzing tree-level
four-point amplitudes relevant to $\gamma$+ jet and dijet final
states.  We make use of four gauge boson amplitudes evaluated near the
first resonant pole to determine the discovery potential of LHC for
$g^*$ and $C^*$ excitations.  We study the inclusive dijet mass
spectrum in the central rapidity region $|y_{\rm jet}| < 1.0$ for
dijet masses $M\geq 2.5~{\rm TeV}$. We find that with an integrated
luminosity of 100~fb$^{-1}$, the 5$\sigma$ discovery reach can be as
high as 4.7~TeV. Observations of resonant structures in $pp\rightarrow
{\rm direct}\ \gamma~ +$ jet can provide interesting corroboration for
string physics up to 3.0~TeV. We also study the ratio of dijet mass
spectra at small and large (center-of-mass) scattering angles. We show
that with the first~fb$^{-1}$ such a ratio can probe lowest-lying
Regge states for masses $\sim 2.5$~TeV. The outline of the paper is as
follows: in Sec.~II we collect all the relevant formulae leading to
four gauge boson string amplitudes, the analysis of the LHC is carried
out in Sec.~III, and we summarize in Sec.~IV.

\section{Four-point amplitudes of gauge bosons}

The most direct way to compute the amplitude for the scattering of
four gauge bosons is to consider the case of polarized particles
because all non-vanishing contributions can be then generated from a
single, maximally helicity violating (MHV), amplitude -- the so-called
{\it partial\/} MHV amplitude~\cite{Parke:1986gb}.  In canonical
D-brane constructions, all string effects are encapsulated in this
amplitude in one ``form factor'' function of Mandelstam variables
$s,~t,~u$ (constrained by $s+t+u=0$):\footnote{For simplicity, we drop
  carets for the parton subprocesses.}
\begin{equation}
V(  s,   t,   u)= \frac{s\,u}{tM_s^2}B(-s/M_s^2,-u/M_s^2)={\Gamma(1-   s/M_s^2)\ \Gamma(1-   u/M_s^2)\over
    \Gamma(1+   t/M_s^2)}.\label{formf}
\end{equation}
The physical content of the form factor becomes clear after using the
well-known expansion in terms of $s$-channel resonances~\cite{Veneziano:1968yb}
\begin{equation}
B(-s/M_s^2,-u/M_s^2)=-\sum_{n=0}^{\infty}\frac{M_s^{2-2n}}{n!}\frac{1}{s-nM_s^2}
\Bigg[\prod_{J=1}^n(u+M^2_sJ)\Bigg],\label{bexp}
\end{equation}
which exhibits $s$-channel poles associated to the propagation of
virtual Regge excitations with masses $\sqrt{n}M_s$. Thus near the
$n$th level pole $(s\to nM^2_s)$:
\begin{equation}\qquad
V(  s,   t,   u)\approx \frac{1}{s-nM^2_s}\times\frac{M_s^{2-2n}}{(n-1)!}\prod_{J=0}^{n-1}(u+M^2_sJ)\ .\label{nthpole}
\end{equation}
In specific amplitudes, the residues combine with the remaining
kinematic factors, reflecting the spin content of particles exchanged
in the $s$-channel, ranging from $J=0$ to $J=n+1$. Unfortunately,
Veneziano amplitudes only apply to strings propagating on flat
Minkowski backgrounds, and their generalization to warped spaces is
presently unknown. In the absence of concrete string theory
constructions, we describe the lowest-lying Regge excitations of SM
gauge bosons following the bottom-up approach advocated
in~\cite{Perelstein:2009qi}.  In the limit where $k$ is taken to zero, this innovative
approach  reproduces the string effects encapsulated in (\ref{formf}).

Consider a free (non-interacting) massive spin-2 field $B_{MN}$ in
curved 5-dimensional spacetime,
\begin{equation}
{\cal L} =  \frac 1 4 H^{LMN}H_{LMN} - \frac 1 2  H^{LM}{}_{M} H_{LN}{}^{N} + \frac{1}{2} m^2 \left[ \left({B_M}^M \right)^2 - B^{MN}B_{MN} \right] \,,
\label{T5dL}
\end{equation}
where $H_{LMN} = \nabla_L B_{MN} - \nabla_M B_{LN}$ is the field
strength tensor and $m \equiv M_s^\star$ is the mass of the lightest Regge
excitation.  This field can be further decomposed according to its
spins ($J= 0,$ $J=1,$ and $J=2$) in 4-dimensions. The tensor, vector,
and scalar components are $B_{\mu \nu},$ $B_{\mu5}$, and $B_{55}$,
respectively. The Lagrangian (\ref{T5dL}) contains terms which mix
these components. Such mixed terms need to be canceled for a
consistent KK decomposition.  As shown
in~\cite{Perelstein:2009qi}, the
action can be factorized as
\begin{equation}
S = S_{J=2} \oplus S_{J=1, J=0} \,,
\end{equation}
where the 5-dimensional Lagrangian for $J=2$ is given by \ba\el{5dLB}
S_{J=2} & = & \int d^5 x \left\{e^{2k |y|}\left[ \frac 1 4
    H^{\xl\mu\nu}H_{\xl\mu\nu} - \frac 1 2 \lb 1 - \frac 2 \xi \rb
    H^{\xl\mu}{}_{\mu} H_{\xl\nu}{}^{\nu}\right] \right.  \nn & +&
\frac 1 2 B_\mu{}^\mu (-\p_y^2 + 4 k^2 + m^2) B_\nu{}^\nu - \frac 1 2
B^{\mu\nu} (-\p_y^2 + 4k^2 + m^2) B_{\mu\nu} \nn & + & \left. 2 k
  \phantom{\frac{1}{1}} \left[\xd(y)-\xd(y-\pi r_c) \right]
  \left[B^{\mu\nu} B_{\mu\nu} - (B_\mu{}^\mu)^2 \right]\right\} \, ,
\ea and $\xi$ is a parameter in the gauge fixing term. The field
$B_{\mu\nu}$ can be decomposed according to its wave function in the
warped dimension,
\begin{equation}
B_{\mu\nu} = \frac 1 {\sqrt{\pi r_c}} \sum_{n = 1}^\infty B_{\mu\nu}^{(n)} \  f^{(n)}(y).
\end{equation}
The equation of motion is,
\begin{equation}
e^{2k|y|} D_{\mu\nu}{}^{\xa \xb} B_{\xa \xb} + \{-\p_y^2 + 4 k^2 + m^2-4 k [\xd(y)-\xd(y-\pi r_c)]\}B_{\mu\nu} =0,
\end{equation}
where $D_{\mu\nu}{}^{\xa \xb}$ is an operator from the first line of \er{5dLB}.
A massless spin-2 field has the equation of motion of $D_{\mu\nu}{}^{\xa \xb} B_{\xa \xb} = 0$. So the masses are given by the eigenvalues of the operator,
\begin{equation}
e^{-2k|y|} \{-\p_y^2 + 4 k^2 + m^2-4 k [\xd(y)-\xd(y-\pi r_c)]\},
\end{equation}
with mode functions $f^{(n)}$ satisfying the following equation,
\begin{equation}
-f^{(n)}{}''+(4k^2 + m^2) f^{(n)} - 4 k \left[\xd(y)-\xd(y-\pi r_c) \right]f^{(n)}
= (\mu^{(n)})^2 \, e^{2k|y|} \, f^{(n)},
\label{eigeneq}
\end{equation}
and associated inner product,
\begin{equation}
\frac{1}{\pi r_c} \int_0^{\pi r_c} dy \, e^{2k|y|} f^{(n)}\, f^{(m)} = \delta^{nm},
\label{innerp}
\end{equation}
from the orthonormal condition. For this choice of $f^{(n)}$, we have
(from the second and the third line of \er{5dLB}), \ba \int d^5x \dots
&= & - \frac 1 2 \, \int d^4 x \, dy \, B^{\mu\nu} \{-\p_y^2 + 4 k^2 +
m^2-4 k [\xd(y)-\xd(y-\pi r_c)]\} B_{\mu\nu} \nn & = & - \frac 1 2 \,
\int d^4 x \, dy \, B^{(m)\mu\nu}(x) \, B^{(n)}_{\mu\nu}(x) \frac 1
{\pi r_c} \,\sum_{n=1}^\infty \sum_{m=1}^\infty (\mu^{(n)})^2 \,
e^{2k|y|} \,f^{(m)}(y) \,f^{(n)}(y) \nn & = & - \frac 1 2 \, \int d^4
x \, \sum_{n=1}^\infty (\mu^{(n)})^2 \, B^{(n)\mu\nu}(x)
B^{(n)}_{\mu\nu}(x),\ea where in the last line, we use
(\ref{innerp}). The integration of $H_{\xl \mu\nu} H^{\xl \mu\nu}$ is
trivial because there is no $y$-derivative. Hence, after the extra
dimension is integrated out, Eq.\er{5dLB} can be reduced to a
4-dimensional Lagrangian of free spin-2 fields (with different masses
$\mu^{(n)}$), \ba\el{4dLB} S_{J=2} & = & \int d^4 x \sum_{n=1}^\infty
\left\{ \frac 1 4 H^{(n)\xl\mu\nu}H^{(n)}_{\xl\mu\nu} - \frac 1 2 \lb
  1 - \frac 2 \xi \rb H^{(n)\xl\mu}{}_{\mu}
  H^{(n)}_{\xl\nu}{}^{\nu}\right. \nn & + & \left. \frac 1 2
  (\mu^{(n)})^2 [B^{(n)}_\mu{}^\mu B^{(n)}_\nu{}^\nu -B^{(n)\mu\nu}
  B^{(n)}_{\mu\nu}] \right\} ,\ea where $\xi \to \infty$ when
computing the scattering amplitude.  The general solution of
(\ref{eigeneq}) is a Bessel function~\cite{Perelstein:2009qi}
\begin{equation}
f^{(n)} (y) = \frac{1}{N} \left[J_\nu\left(\frac{\mu^{(n)}}{\Lambda_{\rm IR}} \, w \right) + c J_{-\nu}\left(\frac{\mu^{(n)}}{\Lambda_{\rm IR}} \, w \right) \right] \,,
\end{equation}
where $N$ is the normalization constant, $c$ is an integration
constant (each of these constants implicitly depends upon the level
$n$),  $\Lambda_{\rm IR} = k e^{-\pi kr_c},$ and $w = e^{k(|y| - \pi
  r_c)}, \, \in [e^{-k\pi r_c},1].$ The order of the
Bessel function is $\nu \equiv \sqrt{4+\mathfrak{m}^2}$, where
$\mathfrak{m}=m/k$ is the string scale in units of the RS
curvature. With appropriate boundary conditions, the masses
$\mu^{(n)}$ and the explicit form of $f^{(n)}$ can be obtained.

We now turn to the discussion of $J=0$. In the effective
  4-dimensional theory there is one real scalar $\Re {\rm e} (\phi)$, which comes
  from the 5-dimensional scalar and couples to the gluon strength
  $F^2$.  In addition, there is one pseudoscalar axion $A_\star^5$,
  which in 4 dimensions couples as $A_\star^5 F \, ^{^{*}\!\!}F$, with
  $^{^{*}\!\!}F = \frac{1}{2} \epsilon^{\mu \nu \rho \sigma} F_{\rho \sigma} .$  This
  pseudoscalar axion comes from the fifth component of a massive
  vector $A_\star$, with coupling $\epsilon_{\mu\nu\rho\sigma 5}
  F^{\mu\nu}F^{\rho\sigma} A_\star^5$. Then, $\Re {\rm e} (\phi)$ and
  $\Im {\rm m}
  (\phi) \equiv A_\star^5$ combine to one
  complex scalar $\phi$ which couples as $\phi
  (F+i\, ^{^{*}\!\!}F)(F+i\, ^{^{*}\!\!}F) + {\rm cc}$; this ensures that $\phi$
  and its complex conjugate $\phi^*$ couple only to the
  $++$ ($--$) helicity combinations, respectively.
  \footnote{We may trace the origin of the $J=0$ contribution to
  components of $B_{MN}$ and other fields of the 10-dimensional theory.
  Instead, we proceed by simply using the correspondence with the tree
  level string theory and identify the vertex function through comparison
  with the tree level $J=0$ pole. As described in the text this has the
  correct helicity structure.} Both the scalar and the pseudoscalar will be
  affected in the same way by warping, because they sit in one SUSY
  multiplet.  Thus, to determine the $J=0$ contribution, we study the
  effect of warping on a dilaton-like scalar a with the coupling $\Re
  {\rm e}(\phi)
  F^2.$
%\footnote{It is possible that there is a way to represent
 %      part of $\phi$ by using $B_{MN},$ but one can always rewrite
  %    the relevant fields (by using some "duality" transformations --
   %   for example 2-index antisymmetric tensor can be dualized to a
    %  pseudoscalar in 4 dimensions) and the final representation will be as
     % given above.}

The Klein-Gordon equation for a scalar $\phi$ in the RS spacetime is
\begin{equation}
\frac 1 {\sqrt{g}}\p_M \sqrt{g} \p^M \phi + m^2 \phi = 0 \,;
\end{equation}
more explicitly, it is,
\be
e^{2k|y|} \p^\mu \p_\mu \phi + \left[- \p_y^2 + 4 k \sy \p_y + m^2 \right] \phi = 0 \, .
\ee
The field $\phi$ can be decomposed according to its wave function in
the warped dimension,
\be
\phi(x,y) = \frac 1 {\sqrt{\pi r_c}} \sum_{n = 1}^\infty \phi^{(n)}(x) \,
h^{(n)}(y).
\ee
One can choose the mode functions $h^{(n)}$ satisfying the following equation,
\be\el{modefun} -h^{(n)}{}''+4 k\, \sy h^{(n)}{}' +  m^2 h^{(n)} = (\mu^{(n)})^2 e^{2k|y|} h^{(n)}.\ee
With a change of variable $x = \frac 1 k e^{k|y|}$, we have
\begin{equation}
\frac {d }{dy} = k x \frac {d }{dx},\quad \frac {d^2} {dy^2} =k^2 x^2
\frac {d^2 }{dy^2} + k^2 x \frac {d }{dx},
\end{equation}
so \er{modefun} can be written as
\be x^2 h^{(n)}{}'' + 3 x h^{(n)}{}' + [(\mu^{(n)})^2 x^2 - {\mathfrak m}^2] (\mu^{(n)}) = 0 .\ee
The solution to this equation is
\be
h^{(n)}(x) = \frac 1 N  \left(\mu^{(n)} x \right)^2 \Big\{J_{\nu}
\left(\mu^{(n)} x \right) + C J_{-\nu} \left(\mu^{(n)} x \right) \Big\} \equiv x^2 \tilde f^{(n)}(x),\ee
where $N$ is a normalization constant and $C$ an integration constant.
For later convenience, we also define
a new function $\tilde f^{(n)}$. The boundary conditions are
\begin{equation}
h^{(n)}{}'(0+) - h^{(n)}{}'(0-) = 0
\end{equation}
and
\begin{equation}
h^{(n)}{}'(-\pi r_c+) - h^{(n)}{}'(\pi r_c-) = 0 ,
\end{equation}
where the prime is the derivative with respect to $y$. As in the case of $B_{\mu\nu}$, the mass $\mu^{(n)}$ is determined from the second boundary condition,
\be
x^2 \tilde f^{(n)}{}'(-\pi r_c+) - 2 x k x \tilde f^{(n)}(-\pi r_c+) - x^2 \tilde f^{(n)}{}'(\pi r_c-) - 2 x k x \tilde f^{(n)}(\pi r_c-) = 0,\ee
or
\be\tilde f^{(n)}{}'(-\pi r_c+) - \tilde f^{(n)}{}'(\pi r_c-) = 4 k \tilde f^{(n)}(\pi r_c),\ee
which is essentially the boundary condition for $B_{\mu\nu}$~\cite{Perelstein:2009qi}. As a result, the mass of $\phi$ is exactly the same as that of $B_{\mu\nu}$. Note that $h^{(n)}(x)$ can be expressed as
\begin{equation}
h^{(n)} = e^{2 k |y|} f^{(n)},
\end{equation}
where $f^{(n)}$ are the mode functions for $B_{\mu\nu}$. So $h^{(n)}(x)$ are normalized as
\be
\frac{1}{\pi r_c} \int_0^{\pi r_c} dy \, e^{-2k|y|} h^{(n)}\, h^{(m)} = \delta^{nm},
\ee
This gives a canonical kinetic term for $\phi^{(n)}$ (because of the
different powers of $e^{2k|y|}$).

In this paper we will restrict our calculations to incoming QCD gluons.
We then obtain the decomposition of the QCD gauge
field. Gauge freedom can be used to set $A_5
=0$~\cite{Davoudiasl:1999tf}. This is consistent with the gauge
invariant equation $\oint dx^5 A_5 =0,$ which results from the
assumption that $A_5$ is a $\set{Z}_2$-odd function of the extra
dimension. In this gauge, the 4-dimensional vector zero-mode has a constant
profile in the bulk,
\begin{equation}
A_\mu(x,y) = \frac 1 {\sqrt{\pi r_c}} A_\mu^{(0)}(x)+\dots \,,
\label{agluon}
\end{equation}
 and
the gluon field strength takes the familiar form $F_{\mu \nu}^a
= \partial_\mu A_\nu^a - \partial_\nu A_\mu^a + g_a f^{abc} A_\mu^b
A_\nu^c,$ with $a = 1, \dots, 8$.

The coupling of the 5-dimensional field $B_{MN}$ to the
gluon is given by
\begin{equation}
S_{ggg^*(C^*)} = \int d^5 x \sqrt{-g} \frac {g_5} {\sqrt 2 M_s^\star} C^{abc}\lb F^{aA C} F_{C}^{bB} - \frac 1 4 F^{a C D}F^b_{C D} g^{AB}\rb B^c_{AB}
\label{gF2}
\end{equation}
where $C^{abc} = 2 [{\rm Tr}(T^a T^b T^c) + {\rm Tr} (T^a T^b T^c)]$
is the color factor, $T^a$ are the generators of the fundamental
representation of $U(3)$ (normalized here according to ${\rm Tr}
(T^aT^b) = \frac{1}{2} \delta^{ab})$, and $F_{A B}^a = \partial_A
A_B^a - \partial_B A_A^a + g_5 f^{abc} A_A^b A_B^c$. Note that the
color indices on the field strength $F$ run from 1 to 8; on the tensor
field $B$, $U(3)$ indices ($c = 0, \dots, 8$) are permitted (with $c =0$
corresponding to the tensor excitation $C^*$).\footnote{As can be
  verified from the 4-point function~\cite{Anchordoqui:2007da} there
  is no coupling $gg \to C$, however the composite nature of $C^*$ and $g^*$
  permits respectively $gg \to C^*$ and $gC \to g^*$ couplings, with color globally preserve.}
Hence, $g_5$ is related to the Yang-Mills QCD coupling $g_a$ according
to $g_5 = g_a \sqrt {\pi r_c}$. The factor $g_5/\sqrt{2} M_s^\star$ is
determined by matching the $gg \to g^* (C^*)$ amplitude to the
s-channel pole term in the string (tree-level)
amplitude~\cite{caveats}. Thus, the 4-dimensional coupling term is
found to be \ba\el{4dintgggstar} {\cal L}_{gg \to g^* (C^*)} &=& \frac
{g^{(0)}} {\sqrt 2 \widetilde M_s} C^{abc} \left[\lb F^{\xa \xg}
  F_{\xg}{}^\xb - \frac 1 4 F^{a\xg \xd}F^b_{\xg \xd} \eta^{\xa\xb}\rb
  B^{c\,(0)}_{\xa \xb} + \frac{1}{2} \left( \phi^{c\,(0)} F^{a\mu
      \nu}F^b_{\mu \nu} \phantom{\frac{1}{2}} \right. \right. \nn & +&
\left. \left. \frac{1}{2} \, \bar \phi^{c\,(0)} F^{a\mu \nu}F^{b\rho
      \sigma} \xe_{\mu\nu\rho\sigma} \rb \right], \ea where
  $\widetilde M_s = e^{-k\pi r_c} M_s^\star \sim 1~{\rm TeV}$ is the
  redshifted string scale, $g^{(0)}$ follows from the integration of
  the zero mode $f^{(0)}(y)$ of $B_{\mu\nu}^{(0)}$, and $\bar
  \phi^{c\, (0)}$ is the zero mode for the imaginary part of the
  complex scalar.  Since each field in (\ref{gF2}) contribute to the
  integration with a factor $(\pi r_c)^{-1/2}$ we obtain,
\begin{equation}
g^{(0)} = \frac {g_a e^{-\pi k r_c}}{\pi r_c} \int_0^{\pi r_c} d y\,
e^{2 k y} \, f^{(0)}(y).
\end{equation}
The coupling \er{4dintgggstar} gives three vertices:
\be\el{gggstarvertex}i\frac {\sqrt 2  \, g^{(0)}} {\widetilde M_s}
C^{abc}\lb \xS^{\xa \xb}- \frac 1 4 \eta^{\xa \xb} \xS_\xg{}^\xg\rb
b_{\xa\xb},\,
 i\frac {g^{(0)}} {\sqrt 2  \, \widetilde M_s} C^{abc} \xS_\mu {}^\mu
 ,\,
i \frac {g^{(0)}}{\sqrt 2 \widetilde M_s} C^{abc} \ 4 \,
\xe_{\mu\nu\rho\sigma} \,
k_1^\mu \, \epsilon_1^\nu \, k_2^\rho \, \epsilon_2^\sigma   \,,
\ee
where  $b_{\xa\xb}$
is a polarization of $B_{\xa\xb}^{c\,(0)}$, $k_i^\mu$ and
$\epsilon_i^\nu$ (with $i =1,2$) are respectively the momentum
and polarization of the incoming gluons, $\xS^{\xa \xb} = (k_1^\xa \xe_1^\xg - k_1^\xg \xe_1^\xa)(k_{2\xg}
\xe_2^\xb - k_2^\xb \xe_{2\xg} ) + (\xa \leftrightarrow \xb)$,
and its trace
$\xS_\xg{}^\xg = 4(k_1 \cdot \xe_2)(k_2 \cdot \xe_1) - 4(\xe_1 \cdot \xe_2)(k_1
\cdot k_2)$~\cite{Cullen:2000ef}. As in the $J=2$ case, the coupling is determined by matching to
the $J=0$ pole term in the tree-level string amplitude.

Finally, we note that the $J=1$ resonant level exists, but is not accessible in purely gluonic
scattering~\cite{Anchordoqui:2008hi}.

The $s$-channel pole terms of the average square amplitudes contributing
to $\gamma$+ jet and  dijet production at the LHC can be obtained from the general
formulae given in Ref.~\cite{Lust:2008qc}. The 4-gluon average square amplitude is
given by
\begin{equation}
|{\cal M} (gg \to gg)| ^2  =  2 \
\left(\frac{g^{(0)}}{\widetilde M_s}\right)^4 \ \left(\frac{N^2-4+(12/N^2)}{N^2-1}\right)
 \ \frac{s^4+  t^4 +   u^4}{(  s - \mu^2)^2} \, ,
\label{ggggpole}
\end{equation}
where to simplify notation we have dropped the
superscript indicating the lowest massive Regge excitation, i.e., $\mu \equiv \mu^{(0)}$.
For phenomenological purposes, the poles
need to be softened to a Breit-Wigner form by obtaining and utilizing
the correct {\em total} widths of the
resonances~\cite{Anchordoqui:2008hi}. After this is done, the
contributions of $gg \to gg$ is as follows:
\begin{eqnarray}
|{\cal M} (gg \to gg)| ^2 & = & \frac{19}{12} \
\left(\frac{g^{(0)}}{\widetilde M_s}\right)^4 \left\{ W_{g^*}^{gg \to gg} \, \left[\frac{s^4}{(  s-\mu^2)^2
+ (\Gamma_{g^*}^{J=0}\ \mu)^2} \right. \right.
\left. +\frac{  t^4+   u^4}{(  s-\mu^{2})^2 + (\Gamma_{g^*}^{J=2}\ \mu)^2}\right] \nonumber \\
   & + &
W_{C^*}^{gg \to gg} \, \left. \left[\frac{s^4}{(  s-\mu^2)^2 + (\Gamma_{C^*}^{J=0}\ \mu)^2} \right.
\left. +\frac{  t^4+  u^4}{(  s-\mu^2)^2 + (\Gamma_{C^*}^{J=2}\ \mu)^2}\right] \right\},
\label{gggg2}
\end{eqnarray}
where
$$\Gamma_{g^*}^{J=0} = 75\,\left(\frac{g^{(0)} \mu}{g_a \widetilde M_s}\right) \, \left(\frac{\mu}{{\rm TeV}} \right)~{\rm GeV}, \quad
\Gamma_{C^*}^{J=0} = 150 \, \left(\frac{g^{(0)} \mu}{g_a \widetilde M_s}\right)
\left(\frac{\mu}{{\rm TeV}}\right)~{\rm GeV},$$
$$\Gamma_{g^*}^{J=2} = 45 \, \left(\frac{g^{(0)} \
    \mu}{g_a \widetilde M_s}\right) \left(\frac{\mu}{{\rm
      TeV}}\right)~{\rm GeV}, \quad \Gamma_{C^*}^{J=2} = 75 \,
\left(\frac{g^{(0)} \mu}{g_a \widetilde M_s}\right)
\left(\frac{\mu}{{\rm TeV}}\right)~{\rm GeV}$$ are the total decay
widths for intermediate states $g^*$, $C^*$ (with angular momentum
$J$)~\cite{Anchordoqui:2008hi,Perelstein:2009qi}. The associated
weights of these intermediate states are given in terms of the
probabilities for the various entrance and exit channels \ba
\el{totalcrossdecom} \frac{N^2-4+12/N^2}{N^2-1} & = & \frac {16}
{(N^2-1)^2}\left[\left(N^2-1\right)\left(\frac{N^2-4}{ 4N}\right)^2+
  \left(\frac{N^2-1}{2N}\right)^2\right]\nn & \propto & \frac {16} {(N^2-1)^2}
\left[(N^2-1)(\xG_{g^*\to gg})^2 + (\xG_{C^*\to gg})^2\right] \,,\ea
yielding
$$
W_{g^*}^{gg \to gg} = \frac{8 (\Gamma_{g^* \to gg})^2}{8(\Gamma_{g^* \to gg})^2 +
(\Gamma_{C^* \to gg})^2} = 0.44, \quad
W_{C^*}^{gg \to gg} = \frac{(\Gamma_{C^*
  \to gg})^2}{8(\Gamma_{g^* \to gg})^2 + (\Gamma_{C^* \to gg})^2} =
0.56 \, , $$
where superscripts $J=2$ are understood to be inserted on all the $\Gamma$'s.

As we pointed out in the Introduction, the hypercharge is a color
composite state containing the photon. The $s$-channel pole term of
the average square amplitude contributing to $gg \to \gamma$ + jet is
given by~\cite{Anchordoqui:2007da}
\begin{eqnarray}
|{\cal M} (gg \to g \gamma)|^2  & = &  \frac{5}{3} \, Q^2 \,
\left(\frac{g^{(0)}}{\widetilde M_s}\right)^4   \Bigg[\frac{s^4}{(  s-\mu^2)^2
+ (\Gamma_{g^*}^{J=0}\ \mu)^2}  +  \left. \frac{ t^4+   u^4}{(  s-\mu^2)^2 + (\Gamma_{g^*}^{J=2}\ \mu)^2}\right]
\label{ggggamma}
\end{eqnarray}
where $Q = \sqrt{1/6} \ \kappa \ \cos \theta_W$ is the product of the
$U(1)$ charge of the fundamental representation ($\sqrt{1/6}$)
followed by successive projections onto the hypercharge ($\kappa$) and
then onto the photon ($\cos \theta_W$).  The $C-Y$ mixing coefficient
is model dependent: in the minimal $U(3) \times Sp(1) \times U(1)$
model it is quite small, around $\kappa \simeq 0.12$ for couplings
evaluated at the $Z$ mass~\cite{Berenstein:2006pk}, which is modestly
enhanced to $\kappa \ \simeq 0.14$ as a result of RG running of the
couplings up to 2.5~TeV~\cite{Anchordoqui:2007da}.  It should be noted
that in models possessing an additional $U(1)$ which partners
$SU(2)_L$ on a $U(2)$ brane~\cite{Antoniadis:2000ena}, the various
assignment of the charges can result in values of $\kappa$ which can
differ considerably from $0.12.$ For the phenomenological analysis
that follows we set $\kappa^2 = 0.02.$

\section{LHC discovery reach}

The most important parameter to determine the LHC discovery reach for
string recurrences is the mass of the lowest-lying Regge excitation,
which depends on $\Lambda_{\rm IR}$ and $\mathfrak{m}$. For fixed
$\mathfrak{m}$ the mass of $g^*$ and $C^*$ excitations is to a very
good approximation a linear function of $\Lambda_{\rm
  IR}$~\cite{Perelstein:2009qi}. As we already remarked in the
Introduction, in Higgsless models $\Lambda_{\rm IR}$ is subject to
significant constraints from electroweak data. The KK excitations of
the vector gauge bosons must be near 1~TeV to simultaneously satisfy
unitarity and electroweak constraints. This leads to $\Lambda_{\rm IR}
\approx 0.5~{\rm TeV}$. Similarly, to avoid precision electroweak
constraints in scenarios where the Higgs is IR-localized the lightest
KK excitation mass is $\agt 3~{\rm TeV}$~\cite{Agashe:2003zs},
yielding $\Lambda_{\rm IR} \agt 1~{\rm TeV}$. From (\ref{Mstar}) we
obtain the condition $\mathfrak{m} \gg 1$ for string propagation on a
smooth geometric background. Nevertheless, as in many examples in
various arenas of physics, $\mathfrak{m} \sim$ a few may in fact be
sufficient, depending on the behavior of the leading corrections to
the geometric limit. In our phenomenological study we will
follow~\cite{Perelstein:2009qi} and set $\mathfrak{m} \agt 3$, which
leads to $\mu^{(0)} \approx 5 \, \Lambda_{\rm IR} $,  $g^{(0)}/g_a
\simeq 0.1,$ and  $\mu^{(0)} = 5 \, \mathfrak{m}^{-1}  \, \widetilde M_s
\simeq 1.7 \widetilde M_s$.

Given the particular nature of the process we are considering, the
production of a TeV Regge state and its subsequent 2-body decay, one
would hope that the resonance would be visible in data binned
according to the invariant mass $M$ of the dijet, after setting cuts
on the different jet rapidities, $|y_1|, \, |y_2| \le
1$~\cite{Bhatti:2008hz} and transverse momenta $p_{\rm T}^{1,2}>50$
GeV.  With the definitions $Y\equiv \frac{1}{2} (y_1 + y_2)$ and $y \equiv
\frac{1}{2} (y_1-y_2)$, the cross section per interval of $M$ for $p
p\rightarrow {\rm dijet}$ is given by
\begin{eqnarray}
\frac{d\sigma}{dM} & = & M\tau\ \sum_{ijkl}\left[
\int_{-Y_{\rm max}}^{0} dY \ f_i (x_a,\, M)  \right. \ f_j (x_b, \,M ) \
\int_{-(y_{\rm max} + Y)}^{y_{\rm max} + Y} dy
\left. \frac{d\sigma}{d\hat t}\right|_{ij\rightarrow kl}\ \frac{1}{\cosh^2
y} \nonumber \\
& + &\int_{0}^{Y_{\rm max}} dY \ f_i (x_a, \, M) \
f_j (x_b, M) \ \int_{-(y_{\rm max} - Y)}^{y_{\rm max} - Y} dy
\left. \left. \frac{d\sigma}{d\hat t}\right|_{ij\rightarrow kl}\
\frac{1}{\cosh^2 y} \right]
\label{longBH}
\end{eqnarray}
where $\tau = M^2/s$, $x_a =
\sqrt{\tau} e^{Y}$,  $x_b = \sqrt{\tau} e^{-Y},$
and
\begin{equation}
  |{\cal M}(ij \to kl) |^2 = 16 \pi \hat s^2 \,
  \left. \frac{d\sigma}{d\hat t} \right|_{ij \to kl} \, .
\end{equation}
In this section we reinstate the caret notation ($\hat s,\ \hat t,\
\hat u$) to specify partonic subprocesses. The $Y$ integration range
in Eq.~(\ref{longBH}), $Y_{\rm max} = {\rm min} \{
\ln(1/\sqrt{\tau}),\ \ y_{\rm max}\}$, comes from requiring $x_a, \,
x_b < 1$ together with the rapidity cuts $y_{\rm min} <|y_1|, \, |y_2|
< y_{\rm max}$. The kinematics of the scattering also provides the
relation $M = 2p_T \cosh y$, which when combined with $p_T = M/2 \
\sin \theta^* = M/2 \sqrt{1-\cos^2 \theta^*},$ yields $\cosh y = (1 -
\cos^2 \theta^*)^{-1/2},$ where $\theta^*$ is the center-of-mass
scattering angle.  Finally, the Mandelstam invariants occurring in the
cross section are given by $\hat s = M^2,$ $\hat t = -\frac{1}{2} M^2\
e^{-y}/ \cosh y,$ and $\hat u = -\frac{1}{2} M^2\ e^{+y}/ \cosh y.$

\begin{figure}
 \postscript{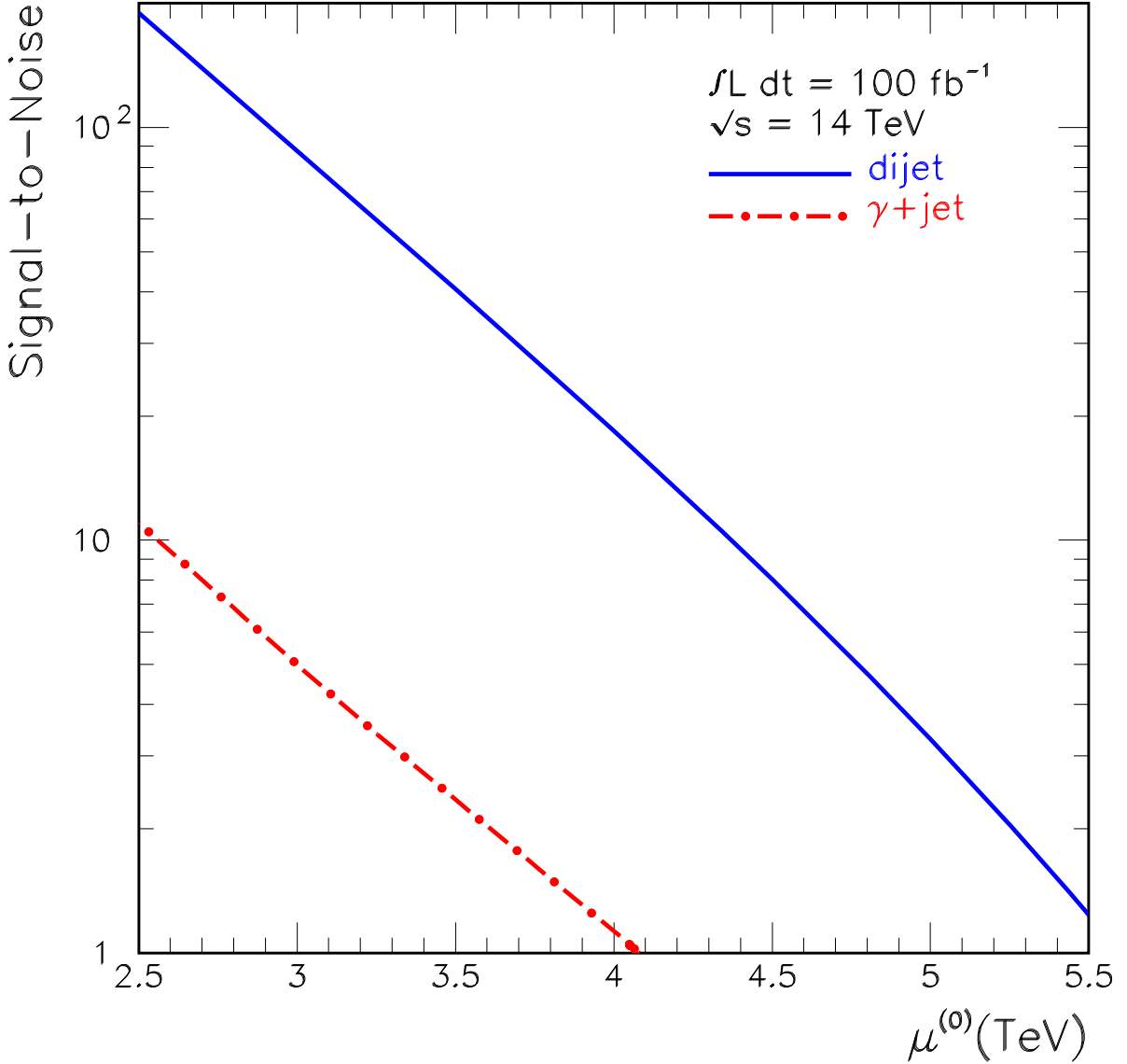}{0.98}
 \caption[$pp \to {\rm dijet}$ signal-to-noise ratio]{$pp \to {\rm
     dijet}$ and $pp \to \gamma + {\rm jet}$ signal-to-noise ratio for
    100~fb$^{-1}$ integrated luminosity.}
\label{RS_S2N}
\end{figure}

Standard bump-hunting
methods, such as calculating cumulative cross sections
\begin{equation}
\sigma (M_0) = \int_{M_0}^\infty  \frac{d\sigma}{dM} \, \, dM
\end{equation}
and searching for regions with significant deviations from the QCD
background, may allow to find an interval of $M$ suspected of
containing a bump.  With the establishment of such a region, one may
calculate a signal-to-noise ratio, with the signal rate estimated
in the invariant mass window $[\mu^{(0)} - 2 \Gamma, \, \mu^{(0)} + 2 \Gamma]$. As
usual, the noise is defined as the square root of the number of
background events in the same dijet  mass interval for the same
integrated luminosity. The QCD background has been calculated
at the partonic level considering all SM contributions to dijet final
states~\cite{Anchordoqui:2008di}.  Our calculation, making use of the
CTEQ6 parton distribution functions~\cite{Pumplin:2002vw} agrees with
that presented in~\cite{Bhatti:2008hz}.

The top curve in Fig.~\ref{RS_S2N} shows the behavior of the
signal-to-noise (S/N) ratio as a function of the lowest massive Regge
excitation, for 100~fb$^{-1}$ of integrated luminosity and $\sqrt{s} =
14$~TeV. {\it Regge
  excitations with masses $\mu^{(0)} \alt 4.7$~TeV are open to discovery at
  the $\geq 5\sigma$ level.} This implies that in the Higgsless model
 discovery would be possible in a wide range of the presently
unconstrained parameter space, whereas in the model with a Higss
localized on the IR-brane the LHC discovery potential would be only marginal.
The bottom curve in Fig.~\ref{RS_S2N} shows the S/N ratio in
the $pp \to$ direct $\gamma$ + jet channel.
 To accommodate the minimal acceptance
cuts on final state photons from the CMS and ATLAS
proposals~\cite{Ball:2007zza}, we set $|y_{\rm max}|<2.4$.
The approximate equality
of the background due to misidentified $\pi^0$'s and the QCD
background~\cite{Gupta:2008zza}, across a range of large $p_T^\gamma$
as implemented in Ref.~\cite{Anchordoqui:2007da}, is maintained as an
approximate equality over a range of $\gamma$-jet invariant masses
with the rapidity cuts imposed.
Observations of resonant structures in $pp\rightarrow {\rm direct}\
\gamma~ +$ jet can provide interesting corroboration for string
physics up to 3.0~TeV. Before proceeding, we stress that the results
shown in Fig.~\ref{RS_S2N} are conservative, in the sense that we have
not included in the signal the stringy contributions of processes
containing fermions. These will be somewhat more model dependent since
they require details of the SM pattern of masses and mixings, but we
expect that these contributions can potentially increase the reach of
LHC for discovery of Regge recurrences.

\begin{figure}[tbp]
\postscript{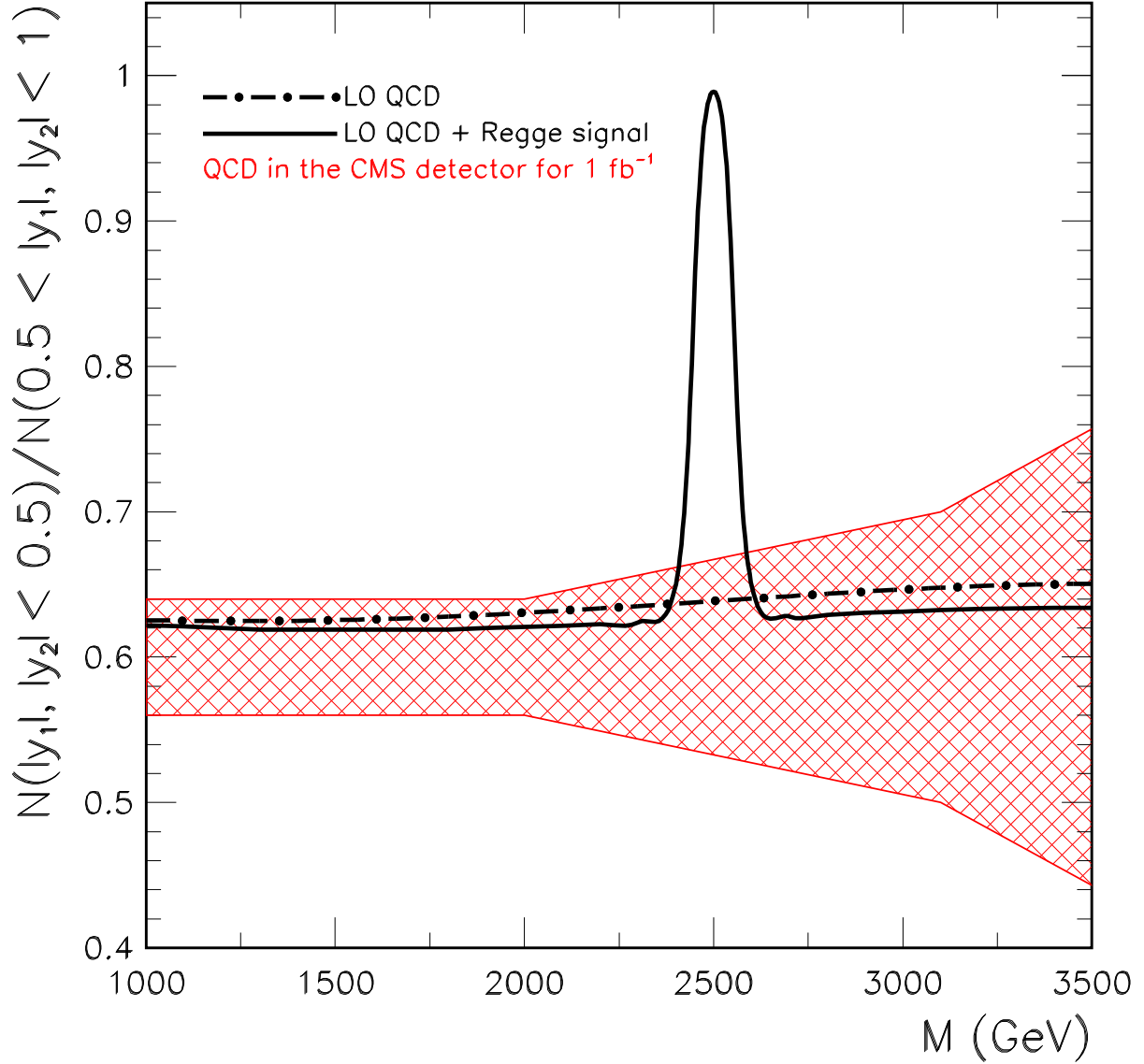}{0.99}
\caption{For a luminosity of 1~fb$^{-1}$, the expected statistical error (shaded region) of the
  dijet ratio of QCD in the CMS detector~\cite{Esen} is compared with
  LO QCD (dot-dashed line) and LO QCD plus lowest massive Regge
  excitation (solid line), for $\mu^{(0)} = 2.5$~TeV.}
\label{RS_R}
\end{figure}

QCD parton-parton cross sections are dominated by $t$-channel
exchanges that produce dijet angular distributions which peak at small
center of mass scattering angles. In contrast,
non--standard contact interactions or excitations of resonances result
in a more isotropic distribution. In terms of rapidity variable for
standard transverse momentum cuts, dijets resulting from QCD processes
will preferentially populate the large rapidity region, while the new
processes generate events more uniformly distributed in the entire
rapidity region. To analyze the details of the rapidity space the D\O\
Collaboration introduced a new parameter~\cite{Abbott:1998wh},
\begin{equation}
R = \frac{d\sigma/dM|_ {(|y_1|,|y_2|< 0.5)}}{d\sigma/dM|_{(0.5 < |y_1|,|y_2| < 1.0)}} \, ,
\end{equation}
the ratio of the number of events, in a given dijet mass bin, for both
rapidities $|y_1|, |y_2| < 0.5$ and both rapidities $0.5 < |y_1|,
|y_2| < 1.0$.  The ratio $R$ is a genuine measure of the most
sensitive part of the angular distribution, providing a single number
that can be measure as a function of the dijet invariant
mass~\cite{Meade:2007sz}.

In Fig.~\ref{RS_R} we compare the results from a full CMS detector
simulation of the ratio $R$, with predictions from LO QCD and
contributions to the $g^*$ and $C^*$ excitations. The synthetic
population was generated with Pythia, passed through the full CMS
detector simulation and reconstructed with the ORCA reconstruction
package~\cite{Esen}. It is clear that with the first~fb$^{-1}$ of data
collected at the LHC, the $R$-parameter will be able to probe
lowest-lying Regge excitations for $\mu^{(0)} \sim
2.5$~TeV.\footnote{It should be noted that the $R$ parameter serves only
  as the crudest discriminator between QCD and stringy behavior of the
  cross section. More detailed analyses of the rapidity dependence of
  the final state jets are in order. In a recent
  paper~\cite{Kitazawa:2010gh} the behavior of the stringy amplitudes
  (for flat geometries) with respect to the rapidity difference $y$
  has been discussed. Results were presented for the separate
  contributions of the 1/2 and 3/2 resonances for the dominant
  $qg\rightarrow qg$ process, as well as for the combined cross
  sections. It remains to compare these to QCD.}

\section{Conclusions}

In this paper, we have extended the work in
Refs.~\cite{Hassanain:2009at} and \cite{Perelstein:2009qi} on an
approximate calculation of string amplitudes in the RS geometry to
include the $J=0$ contribution to bosonic 4-point functions. We have
carried out a phenomenological analysis of the resonant contributions
to dijet production at the LHC, and found that for an integrated
luminosity of 100~fb$^{-1}$, discovery of the resonant signal at
signal-to-noise of 5$\sigma$ is possible for resonant masses of up to
nearly 5~TeV.  However, it should be noted that this is possible only
for the Higgsless model: For the model with the Higgs on or near the
IR brane, the requirement $\Lambda_{\rm IR}\ge $ 1~TeV combined with
the relation $\mu\simeq 5\Lambda_{\rm IR}$ implies $\mu> $ 5~TeV,
greatly narrowing the possible region of discovery.

In addition to the Regge recurrences there are of course KK modes of
SM particles and gravitons propagating in the $s$-channel, which at
this point we have omitted consideration. Their importance can be
gauged by their masses relative to $\mu^{(0)}$.  The ratio of string
to KK masses is model dependent, but in general there could be several
cases where the $\mu^{(0)}/m_{\rm KK}$ ratio is around a
few~\cite{Reece:2010xj}. This relation can be illustrated by comparing
with the masses of the KK states of the graviton: $m_{G}^{(n)} = x_n
\Lambda_{\rm IR}$, with $x_n$ being the $n^{\rm th}$ roots of the
Bessel function $J_1$~\cite{Davoudiasl:1999jd}.  We find that
$\mu^{(0)}/m_{ G}^{(1)} \sim 1.25$. This implies that the KK
contribution is not significantly enhanced over the Regge
contribution~\cite{Hassanain:2009at}, and so here we have limitted our
discussion to the Regge case.

Finally, the large amount of data required for discovery may be traced to
a strong difference at the phenomenological level between the
RS scenario and the flat space result: the effective 4D coupling constant $g^{(0)}\simeq 0.1\ g_a.$
For a given resonance mass, we also have $\widetilde{M}_s\simeq 0.6\mu.$ The net result, following
from Eq.(\ref{gggg2}) is that
for a given resonance mass, the RS cross section is a factor of $(0.1/0.6)^4\approx 10^{-3}$ times
that of the flat case scenario. (There is also some effect from the narrowing of the total widths.)
The drastic reduction of the effective coupling is a direct result
of permitting the gluon field to propagate in the warped bulk.

\section*{Acknowledgements}

We would like to thank Dieter L\"ust for a careful reading of the
manuscript and helpful comments. L.A.A.\ is supported by the
U.S. National Science Foundation (NSF) Grant No PHY-0757598, and the
UWM Research Growth Initiative.  H.G.\ is supported by the NSF Grant
No PHY-0757959.  The research of T.R.T.\ is supported by the U.S. NSF
Grants PHY-0600304, PHY-0757959.  Any opinions, findings, and
conclusions or recommendations expressed in this material are those of
the authors and do not necessarily reflect the views of the National
Science Foundation.

\end{document}